\documentclass[aps,nofootinbib,floatfix,showpacs,preprintnumbers]{revtex4} 

\newcommand{\sfig}[2]{
\includegraphics[width=#2]{#1}
        }
\newcommand{\Sfig}[2]{
    \begin{figure}[thbp]
    \sfig{#1.eps}{.7\columnwidth}
    \caption{{\small #2}}
    \label{fig:#1}
    \end{figure}
}

\newcommand{\rf}[1]{\ref{fig:#1}}

\def\be{\begin{equation}}

\newcommand{\vs}{\nonumber\\}
\newcommand{\ec}[1]{Eq.~(\ref{eq:#1})}

\newcommand{\eql}[1]{\label{eq:#1}}
\usepackage{graphicx}
\usepackage{amsmath}
\usepackage{amsfonts}
\usepackage{amssymb}
\usepackage{latexsym}
\usepackage{color}
\newcommand\ee{\end{equation}}
\newcommand\eea{\end{eqnarray}}
\newcommand\bea{\begin{eqnarray}}
\newcommand\nn{\nonumber}

\renewcommand\({\left(}
\renewcommand\){\right)}
\renewcommand\[{\left[}
\renewcommand\]{\right]}
\newcommand\lsim{\mathrel{\rlap{\lower4pt\hbox{\hskip1pt$\sim$}}
    \raise1pt\hbox{$<$}}}
\newcommand\gsim{\mathrel{\rlap{\lower4pt\hbox{\hskip1pt$\sim$}}
    \raise1pt\hbox{$>$}}}

\newcommand{\newc}{\newcommand}

\newc{\mhalf}{m_{1/2}}      \newc{\mzero}{m_0}

\newc{\tanb}{\tan\beta}
\newc{\azero}{A_0}
\newc{\at}{A_t} \newc{\abot}{A_b} \newc{\atau}{A_\tau} 

\newc{\bmu}{B\mu}           \newc{\sgn}{{\rm sgn}}
\newc{\mone}{M_1}           \newc{\mtwo}{M_2}

\newc{\charone}{\chi_1^\pm} \newc{\mcharone}{m_{\chi_1^\pm}}

\newc{\mgut}{M_{\rm GUT}}

\newc{\mplanck}{M_{\rm Pl}}      
\newc{\mpl}{M_{\rm Pl}}
\newc{\msusy}{M_{\rm SUSY}}      \newc{\ms}{M_{\rm S}}
\newc{\xf}{x_f}
\newc\vrel{v_{\rm rel}}

\newc{\abund}{\Omega h^2}
\newc{\abundchi}{\Omega_\chi h^2}
\newc{\abundcdm}{\Omega_{{\rm CDM}} h^2}

\newc{\omeganlsp}{\Omega_{{\rm NLSP}}}  \newc{\abundnlsp}{\Omega_{\rm NLSP}h^2}
\newc{\omegalsp}{\Omega_{{\rm LSP}}}  \newc{\abundlsp}{\Omega_{\rm LSP}h^2}
\newc{\omegawmap}{\Omega_{{\rm WMAP}}}\newc{\abundwmap}{\Omega_{\rm WMAP}h^2}
\newc{\omegagravitinio}{\Omega_{{\gravitino}}}
\newc{\abundgravitino}{\Omega_{\gravitino}h^2}

\newcommand\gev{\,\mbox{GeV}\ }

\newcommand\kev{\,\mbox{keV}\ }

\newc\gbar{{\overline{g}}}

\newcommand\cm{\,\mbox{cm}\ }
\renewcommand{\newc}{\newcommand}

\newc{\ynlsp}{Y_{{\rm NLSP}}}            \newc{\taunlsp}{\tau_{{\rm NLSP}}}
\newc{\nnlsp}{n_{{\rm NLSP}}}            \newc{\mnlsp}{m_{{\rm NLSP}}}
\newc{\mlsp}{m_{{\rm LSP}}}
\newc{\ylsp}{Y_{{\rm LSP}}}        
\newc{\nx}{n_{X}}                        \newc{\yx}{Y_{X}}
\newc{\mx}{m_{X}}                        \newc{\taux}{\tau_{X}}
\renewcommand{\d}{\textrm{d}}

\newc{\ra}{\rightarrow}

\newcommand\gravitino{\widetilde{G}}    

\newcommand{\bfrac}[2]{{\left(\frac{#1}{#2} \right)  }}

\newc{\gstar}{g_\ast}           \newc{\gsstar}{g_{s\ast}}

       \def\pslash{\not{\hbox{\kern-2.3pt $p$}}}
       \def\kslash{\not{\hbox{\kern-2.3pt $k$}}}
       \def\qslash{\not{\hbox{\kern-2.3pt $q$}}}
       \def\ddslash{\not{\hbox{\kern-2.3pt $d$}}}
       \def\prtslash{\not{\hbox{\kern-2.3pt $\partial$}}}

\begin{document}

\title{Cold Positrons from Decaying Dark Matter}

\author{Lotfi Boubekeur$^{1.2}$, Scott Dodelson$^{3,4,5}$, Oscar Vives$^{1,2}$}

\affiliation{$^1$ Departament de F\'{\i}sica Te\`orica, Universitat de Val\`encia, E-46100, Burjassot, Spain}
\affiliation{$^2$ Instituto de F\'{\i}sica Corpuscular (IFIC), Universitat de Val\`encia-CSIC,
Edicio de Institutos de Paterna, Apt. 22085, E-46071, Valencia, Spain}
\affiliation{$^3$Center for Particle Astrophysics, Fermi National Accelerator Laboratory, Batavia, IL~~60510}
\affiliation{$^4$Department of Astronomy \& Astrophysics, The University of Chicago, Chicago, IL~~60637}
\affiliation{$^5$Kavli Institute for Cosmological Physics, Chicago, IL~~60637}

\begin{abstract}
Many models of dark matter contain more than one new particle beyond those in the Standard Model. Often heavier particles decay into the lightest dark matter particle as the Universe evolves. Here we explore the possibilities that arise if one of the products in a (Heavy Particle) $\rightarrow$ (Dark Matter) decay is a positron, and the lifetime is shorter than the age of the Universe. The positrons cool down by scattering off the cosmic microwave background and eventually annihilate when they fall into Galactic potential wells. The resulting 511 keV flux not only places constraints on this class of models but might even be consistent with that observed by the INTEGRAL satellite.
\end{abstract}
\preprint{IFIC/12-41, FTUV-12-0611}
\maketitle

\section{Introduction}

Although there is ample evidence for the existence of non-baryonic dark matter, the properties of the particle[s] that make up the dark matter are not well determined. This is not surprising as the evidence for dark matter to date is from observations of its gravitational effects. But this situation may change soon as more powerful direct and indirect detection experiments come online; indeed, there are already numerous hints from both sectors. Direct detection experiments DAMA/LIBRA \cite{Bernabei:2010mq}, COGENT \cite{Aalseth:2010vx}, CRESST \cite{Angloher:2011uu} have events consistent with a dark matter signal and there are several hints of indirect detection from the Fermi Gamma Ray Satellite and radio observations~\cite{Hooper:2010mq,Linden:2011au}. There are also two sets of observations of positrons that could be explained by dark matter: the observed excess of positrons over electrons in PAMELA \cite{Adriani:2008zr} and INTEGRAL observations of the 511\kev line from the centre of the galaxy (see, e.g., \cite{Jean:2003ci,Knodlseder:2003sv,Boehm:2003bt} and \cite{Prantzos:2010wi} for a review).

It has been suggested~\cite{Hooper:2004qf,Picciotto:2004rp,Pospelov:2007xh,Cembranos:2007vk} that the INTEGRAL observations can be explained by dark matter decays\footnote{See, however, Ref.~\cite{Lingenfelter:2009kx} for problems with the dark matter interpretation. While the model proposed here circumvents some of these (the positrons being quite cold as they enter the disk and bulge), tension remains in the bulge to disk ratio.}. The observed flux of photons $F_{\rm INTEGRAL}\simeq 10^{-3}$ ph cm$^{-2}$ sec$^{-1}$ is apparently produced by the annihilation of positrons with galactic electrons almost at rest.  Known astrophysical sources cannot account for the totality of these positrons \cite{Prantzos:2010wi}, so it is natural to consider positrons produced by dark matter annihilations or decays. In the decay scenarios considered so far, a small fraction $(t_0/\tau_{\rm DM})$ of dark matter particles decay at the present time into low momenta positrons ($\sim$few MeV) in the bulge, and the positrons subsequently annihilate with electrons to produce the 511 keV line. The required lifetime is larger than the age of the universe $\tau_{\rm DM}\sim 10^{20}\sec \(100\, \gev/m_{\rm DM}\)$.

Here we explore the possibility that the required positrons were produced at earlier times. For concreteness, we assume that the sector containing dark matter has a stable component $\chi$ (which is the dark matter today) and an unstable component, $X$, that decays into $\chi$ and positrons some time after recombination. Even within this simple scenario, there are a number of dials to turn: the masses of $\chi$ and $X$; the lifetime of $X$; the charge of $X$; the relative abundance of $X$ (since $\chi$ is the dark matter today, its abundance is fixed by observations); and the branching ratios for the $X$ decay into positrons and photons. What happens to the decay-produced positrons depends on the values of these parameters, and we find a rich spectrum of observational consequences. Thus, although motivated by the 511 keV excess, we will analyze the constraints and possibilities of the generic idea of a heavy particle decaying into dark matter and positrons.

The main difference between our proposal and others is that the positrons in this class of models cool in the early universe very efficiently before they can annihilate. Therefore, when they enter our Galaxy, they are already very cold. This contrasts with models in which the positrons are produced in decays or annihilations of Galactic dark matter, typically concentrated near the center of the Galaxy. In those models, cooling the positrons before they can annihilate is a central problem.

Section II introduces notation, the constraints, and physical processes that are relevant for any positron scenario. The next sections elaborate on the details of cosmological down-scattering (III); energy loss in our Galaxy (IV); and capture in the bulge (V). We conclude in \S VI with a discussion of viable particle physics models.

\section{General Considerations and Overview of Constraints}

We consider the situation where some fraction of the dark matter in the universe is unstable. The decaying particle is called $X$ and it decays to $\chi$, stable dark matter, together with other decay products. We parametrize the number density of $X$-particles by its value early on (say at recombination) before decays start. Define
\be 
\omega \equiv  \frac{n_X}{n_{\chi}}\vert_{z=z_{\rm rec}}
.\ee 
From CMB observations, we know that the dark matter density has been approximately constant (apart from the expansion) since recombination. In our scenario, this is still possible if $\omega< 1$, for 
then the decays produce little change in the dark matter density and we can safely set $n_\chi(z)=n_\chi(z=0)(1+z)^3$. The number density of $X$ as a function of time is therefore
\bea 
n_X &=& \omega n_\chi e^{-t/\tau}
\vs 
&=&
1.2\times 10^{-8}\,{\rm cm}^{-3} (1+z)^3  \omega  \frac{100 \, {\rm GeV}}{m_\chi} e^{-t/\tau}
\eql{defom}
\eea
where $\tau$ is the lifetime of $X$ and we have adopted the WMAP value of $\Omega_{\rm cdm}h^2=0.11$. Most of the constraints (and possibilities) will depend on the branching ratios, in particular what fraction of the decays produce positrons and photons. In this work we analyze the physics of relic positrons, therefore we will focus on the case where the dominant decay mode includes a positron\footnote{The simplest way to ensure this is to require a small mass difference, $\Delta M < 2$ GeV for neutral $X$ or  $\Delta M < 1$ GeV for charged $X$. With this small mass difference, the decay to protons is kinematically forbidden and all decays produce electrons/positrons (with the exception of neutral pions that produce photons). A more general model with larger mass differences might also work but the constraints from the decay-produced protons are more complex than those we consider here.}.  The decay to photons is nevertheless very important due to the stringent constraints on diffuse gamma rays. In general, we take the branching ration to photons, $B_\gamma$, as a free parameter. In some cases $B_\gamma$ can be of order one. An example of this is the neutralino decay into photon plus gravitino in MSSM theories with gravitino LSP. In any case, the expected value of $B_\gamma$ is at least of order $\alpha$, since the Feynman diagram of the dominant mode can be extended to have the positron emit a photon, leading to a decay rate a factor of $\alpha$ smaller.

There is also the question of the charge of the decaying particles. The simplest case regarding the phenomenological constraints is if $X$ is neutral. If $X$ is charged, the relic abundances of $X^+$ and $X^-$ are likely the same, and we call this the symmetric scenario.  But there is also the possibility of an asymmetry in $X^+$ versus $X^-$ (compensated by a net charge in ordinary particles so the Universe remains electrically neutral) similar to the observed particle anti-particle asymmetry. The constraints differ in all three of these cases (neutral, charged symmetric and charged antisymmetric), so we consider them separately.

There are five sets of constraints on this scenario:
\begin{enumerate}
  \item {\bf Collider constraints.} If $X$ is charged and light enough, it can be produced in colliders and will leave a distinctive signature. 
   \item {\bf Catalyzed Big Bang Nucleosynthesis (CBBN).} If $X$ is charged and the abundance is symmetric, the $X^-$ would bind to light nuclei and catalyze nucleosynthesis, violating the success of the comparison with observed light element abundances today.
  \item {\bf Heavy Water.} If $X$ is charged both in the symmetric and asymmetric cases, the remaining $X^+$ today could bind with an electron and $X^-$ with $^4$He and produce heavy water, the abundance of which is tightly constrained.
  \item {\bf Diffuse Photon Flux from Direct Decays.} A fraction $B_\gamma$ of decays will produce high energy photons. Photons produced at $z \lsim 1000$ with energies today between $100$ keV and $100$ GeV travel freely through the Universe (see, e.g., \cite{zdziarski89,Chen:2003gz,Cirelli:2010xx} and Fig.~\ref{figOD} below) so would be observed as part of the diffuse X-ray or Gamma-ray background today.
  \item {\bf Diffuse Photon Flux from Inverse Compton Scattering.} Even if $B_\gamma$ were zero, positrons (and electrons) produced in decays produce high energy photons via Inverse Compton Scattering off the Cosmic Microwave Background. These would also be part of the diffuse background today except for energies below $100$ keV and large $z$ where the absorption by the intergalactic medium is important.
 \end{enumerate}
These constraints, and the scenario to which they apply, are summarized in Table I.

\begin{table}[htbp]
\begin{center}
\begin{tabular}{|l||c |c |c |}
\hline
Constraint & Neutral $X$ & Charged $X$/Symmetric & Charged $X$/Asymmetric\\
\hline\hline
Collider & No & Yes & Yes \\
\hline
CBBN & No & Yes & No \\
\hline
Heavy Water & No & Yes & Yes \\
\hline
Diffuse Flux: Direct Decays & Yes & Yes & Yes \\
\hline
Diffuse Flux: Inverse Compton & Yes & Yes & Yes \\
\hline
\end{tabular}
\caption{Constraints on three scenarios of decay-produced positrons. {\it Asymmetric} means the relic heavy $X$ particles are positively charged.} 
\end{center}
\vspace{-0.6cm}
\end{table}

Items 1, 2, and 3 have been studied carefully in the literature so we can lift results from previous work. 

Because of the presence of electromagnetic couplings, one can hope to produce these heavy long-lived charged particles through the well-known electroweak processes. Popular examples of such particles are long-lived staus and gluinos $R$-hadrons in the MSSM. The LHC with its two dedicated detectors is probing these candidates. The ATLAS collaboration \cite{Aad:2011hz} obtained a model-independent bound using 37 pb$^{-1}$ of data at center-of-mass energy $\sqrt{s}=7$ TeV. This bound can be applied to the mass of the charged NLSP giving $m_{X^\pm}\ge 110$ GeV. The CMS collaboration \cite{Chatrchyan:2012sp}, with its 5 pb$^{-1}$ of collected data, is able to put an even stronger bound on the mass $m_{X^\pm}\ge 223$ GeV.

For the neutral case, for instance in $e^+ e^-$ colliders like LEP, one way to look for DM is look for mono-photon events:  $e^+ e^-\to E_T\!\!\!\!\!/+\gamma$. These events are produced through the same interaction responsible for decays,  with one photon attached to $e^\pm$. More precisely the reaction is   $e^+ e^-\to \chi\, X +\gamma$, where $\chi$ and $X$ appear as missing energy.   This analysis has been applied in the past to a variety of scenarios (see e.g.\cite{Fox:2011fx} for LEP). Unfortunately in our case, this strategy does not lead to a significant constraint because the operators that lead to decays predict very small production rates. This conclusion holds also for hadronic machines\footnote{Notice that even though there is no coupling of dark matter to quarks, it will be generated at 1-loop level (See e.g \cite{Kopp:2009et}). }, like the Tevatron and the LHC where one looks for missing transverse energy accompanied by monojets  $pp (p\bar{p})\to E_T\!\!\!\!\!/$ +mono-jet. See e.g. \cite{Bai:2010hh} for analysis on Tevatron data.

In the charged symmetric case, the abundance at BBN should satisfy the (CBBN) constraints, which reads
\be
\omega 
\lesssim 
2.44\times10^{-3}\bfrac{m_X}{100\gev}\bfrac{Y_{\textrm{CBBN}}}{10^{-14}}
\label{CBBNomega}
\ee
where $Y_{\textrm{CBBN}}$ is the maximum value allowed from catalyzed
BBN for $X^-$-lifetimes larger than $10^5\sec$. There are two possible
values for $Y_{\rm CBBN}$ available in the literature: the
conservative one \cite{Jedamzik:2007qk,Jedamzik:2009uy} $Y_{\rm
  CBBN}=10^{-14}-10^{-15}$ and the more stringent one
\cite{Hamaguchi:2007mp,Pradler:2007is} $Y_{\rm CBBN}=10^{-16}$.
In both charged scenarios, charged relics $X^\pm$ can form heavy Hydrogen atoms that can condense in 
the form of anomalously heavy water in the bottom of oceans. In this case, 
the so-called heavy water bound applies to the present abundance of heavy
$X^\pm$-particles \cite{Yamagata:1993jq} 
\be
n_{X^\pm}(t_0) < 2.63\times 10^{-35} \,{\rm cm}^{-3}
\label{hw}
\ee
This bound implies that the lifetime of $X^\pm$ must be short enough so that, at present, almost all the $X^\pm$ have already decayed. 
These bounds on charged $X$'s are shown in the $\omega,\tau$ plane in Fig.~\rf{hw}. Note that the Heavy Water constraint requires lifetimes less than $\sim8\times10^{15}$s, corresponds to redshift $z\simeq 15$.

\Sfig{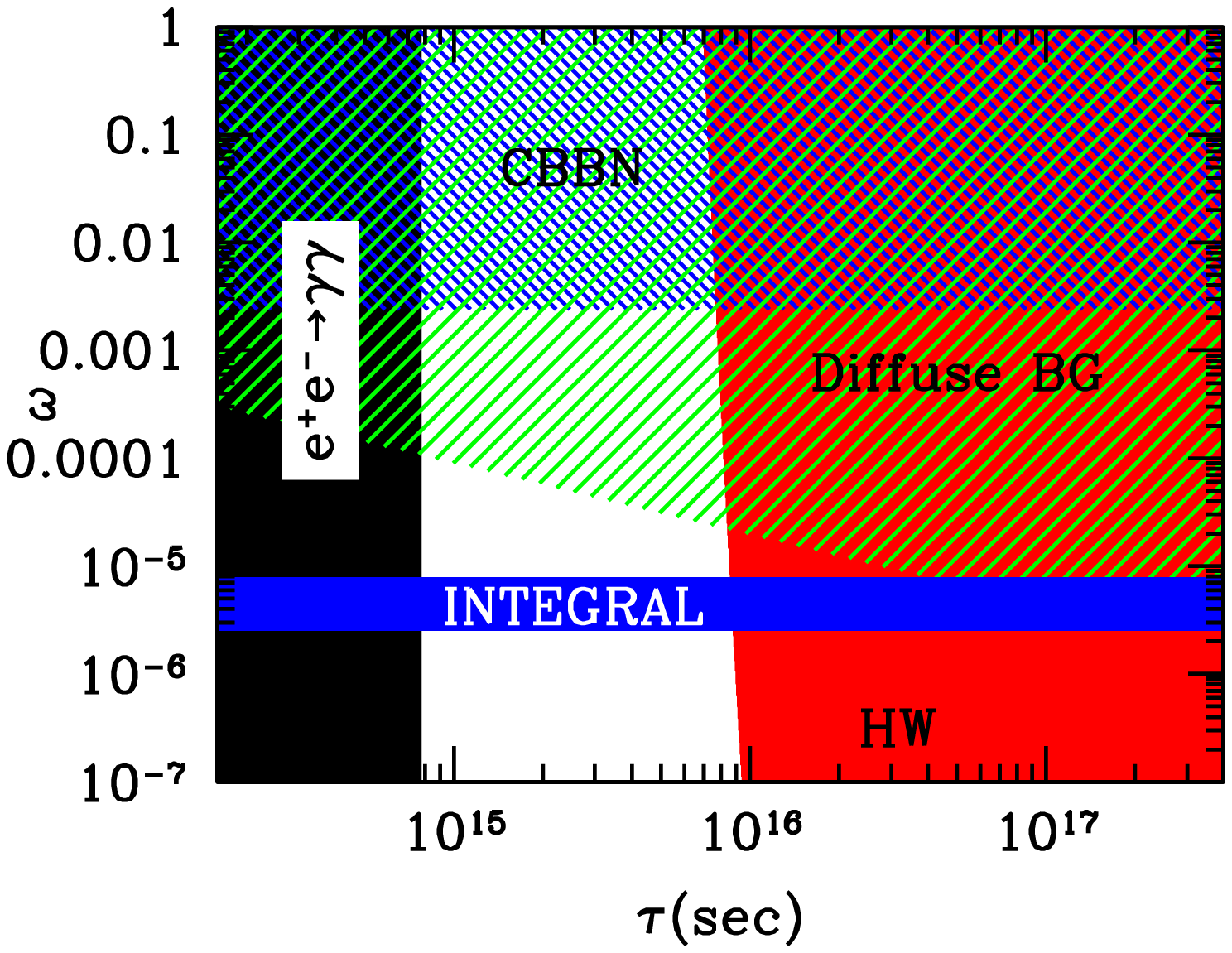}{Constraints on charged $X$'s from Heavy Water (HW), CBBN (with the conservative upper limit $Y_{\rm CBBN}=10^{-14}$) and the diffuse gamma ray background. The CBBN constraint applies only to the symmetric scenario.  The region labelled $e^+e^-\rightarrow \gamma\gamma$ would produce positrons early enough that the high background electron density would lead to complete annihilation and therefore no residual signal. Shaded regions are excluded, and the constraints use  $m_\chi=100$ GeV. The horizontal blue band denotes the region in parameters space that would produce a 511 keV flux consistent with that observed by INTEGRAL. }

\newcommand\deltam{E_{\rm init}}

The diffuse photon flux from the decays is tightly constrained by observations, a constraint that applies to all scenarios.
Photons can be produced directly in decays or indirectly when electrons and positrons scatter down off of the cosmic microwave background. 
We work out the details in Appendices, but here provide estimates for the flux due to inverse Compton scattering. (As shown in the Appendix, the constraints from direct decays are similar.) If the decay-produced positron has initial energy of order $\deltam$  (typically of order $\Delta M/2$)
immediately after decay at time $t=\tau$, then the up-scattered photons will have energies of order $4T(\tau)(\deltam/m_e)^2$ where $T$ is the temperature of the CMB. These photons today will have energy $\sim4T(\tau)(\deltam/m_e)^2/(1+z_d) = 4T_0(\deltam/m_e)^2$ where $z_d$ is the redshift when $t=\tau$. Therefore the diffuse flux will peak at
\be 
E_{\rm max} \sim 3.6 \, {\rm keV} \left(\frac{\deltam}{1\,{\rm GeV}}\right)^2
\ee
independent of the lifetime of $X$ and cut off sharply at higher energies.
The diffuse flux at this peak will roughly be of order
$cn_X(\tau)/(4\pi E)$, if these photons travel freely and are observed today. The more careful calculation in the Appendix, including also the attenuation through Compton scattering on cold electrons and photoionization, leads to
\be 
F \simeq 10^{5} \, {\rm cm}^{-2}\, {\rm s}^{-1}\, {\rm sr}^{-1}\, {\rm keV}^{-1} 
\omega \left( \frac{100\, {\rm GeV}}{m_\chi} \right) \left( \frac{\tau}{10^{15}\,{\rm sec}}\right)^{2/3}
\left( \frac{1\,{\rm keV}}{E} \right)^{3/2}\qquad (\tau<10^{17}\,{\rm sec})
\eql{textflux}
.\ee
Applying the SPI constraint\footnote{{\tt http://heasarc.gsfc.nasa.gov/docs/objects/background/diffuse\_spectrum.html}} at $E=1$ keV, $F < 10\,  {\rm cm}^{-2}{\rm s}^{-1}{\rm sr}^{-1} {\rm keV}^{-1}$ leads to
\be 
\omega \left( \frac{100\, {\rm GeV}}{m_\chi} \right) \left( \frac{\tau}{10^{15}\,{\rm sec}}\right)^{2/3}
< 10^{-4} \qquad (\tau<10^{17}\,{\rm sec})
\ee
The full constraint (extended to larger lifetimes) is depicted in Fig.~\rf{hw} when $m_\chi=100$ GeV and $\Delta M=1$ GeV, and this holds as long as the cut-off $E_{\rm max}$
is above 1 keV, or the initial positron energy is above $0.5$ GeV. The constraint loosens if $\deltam$ is smaller, so that the
flux is cut-off well below a keV.



What about the positrons themselves?
 The energetic positrons almost immediately inverse Compton scatter down off the photons in the cosmic background to become non-relativistic. This
down-scattering continues thereafter, so early decays ($z_d<20$) produced
positrons that thermalized with electrons and protons. These positrons
behave as CHAMPs~\cite{DeRujula:1989fe,Chuzhoy:2008zy} (albeit not the dominant component of the dark matter,
as in the traditional CHAMP scenario), collapsing with baryons and dark
matter to form galaxies. 
As we will see below, they quickly annihilate in the disk or bulge of the Galaxy on timescales much too
short to explain 511 keV radiation in our Galaxy. 
The contribution to the 511 keV flux then is from the
flux of positrons on highly elliptical orbits entering the disk or bulge for the first time now. 
%
In the ensuing sections, we estimate the 511 keV flux from these positrons, subject to the constraints depicted in Fig.~\rf{hw}.
Our conclusion is that -- in the regime of parameter space indicated by the blue line in Fig.~\rf{hw} -- positrons could produce flux with the same amplitude as that seen by INTEGRAL. The corollary to this is that the region above the blue band ($\omega>10^{-5}$) is ruled out by the 511 keV line measurements.

\section{Positron energy loss after decay}

Upon production in decays, 
positrons/electrons lose energy mainly through scattering on CMB photons (Inverse Compton scattering). Taking into account expansion, energy loss is governed by \cite{1992hea..book.....L}
\be
\frac{\d E_{e^+}}{dt}+ {p^2_{e^+} H \over E_{e^+}}=-\frac43 \sigma_T \,\rho_\gamma\, c \(p_{e^+} \over m_e\)^2\, , 
\label{elosst}
\ee
where $\sigma_T$ is the Thompson cross section, $H$ the Hubble expansion rate and $\rho_\gamma$ is the energy density of CMB photons. This expression is valid as long as the center of mass energy squared, which is of order $E_{e} T$, is smaller than $m_e^2$ or equivalently $E_e\ll 10^6\textrm{GeV}/(1+z)$, a criterion that is satisfied over the full range of parameter space we are considering. Using the auxiliary variable $\epsilon\equiv {p_{e^+}}/m_e (1+z)$, we can write Eq.~(\ref{elosst}) in the more compact form
\be 
{\d \ln \epsilon\over d\ln(1+z)}= \Gamma \, (1+z)^{5/2}\, \sqrt{\epsilon^2(1+z)^2+1}\, \quad \textrm{with}\;\;   \Gamma\equiv \frac43\, \sigma_T{ \rho_{\gamma,0} \over H_0\, \Omega_m^{1/2}\,  m_e} \simeq 0.0113 .
\eql{eloss1}
\ee

Although the full result must be obtained through numerical
integration of the above equation from the decay redshift $z_d$ to
today, we can understand the behavior of \ec{eloss1} using some
approximations. First, note
that the final momentum depends only on the redshift of decay $z_d$, but not on the initial positron energy $\deltam$. 
To see this qualitatively, note that in the parameter range of interest, $\epsilon$ starts out large, since $p/m_e$ is initially $\deltam/m_e\sim 10^3$, and we are considering decays after $z\sim100$. The scattering term in \ec{eloss1} is then of order $\Gamma\epsilon (1+z)^{7/2}$, typically very large, corresponding to rapid energy loss. In a very small redshift interval after production, then, $\epsilon$ drops quickly until $\epsilon (1+z)$ falls below one, and the square root in \ec{eloss1} can then be approximated as unity. At that point, the positron momentum is approximately equal to $m_e (1+z)\epsilon = m_e$. That is, inverse Compton scattering almost instantaneously slows down the positrons so that they are non-relativistic.


After this steep drop, $\epsilon(1+z)$ falls below unity, and the square root in \ec{eloss1} reduces to one. Integrating under this approximation leads to a slower but still steady decline in momentum, $\epsilon\propto \exp\{ -(2/5) \Gamma [(1+z_d)^{5/2} - (1+z)^{5/2}] \}$. The earlier the positron is produced the more effective the loss process is, so the final positron momentum decreases with increasing $z_d$. Today, this dependence scales as $e^{-2\Gamma(1+z_d)^{5/2}/5}$. 

Numerical integration exhibits these qualitative features, as shown in Fig.~\rf{int}. There is the steep drop immediately after production, leading to positrons with momentum of order an MeV. Then the energy loss continues, all of which is independent of the initial energy. The earlier the positrons are produced, the longer the losses continue, so the final momentum is smallest for positrons produced earliest. As expected, positrons produced at $z=20$ have  momentum reduced by a factor of about 1000 more (due to the exponential dependence on $z_d$) than those produced at $z=10$. Our loss equation is valid only when the positron energy is much greater than the CMB temperature. So, for large $z_d$, the final
momentum we obtain from this equation tends to zero means that the 
positrons equilibrate with the photons and $\langle p_e\rangle \simeq \sqrt{mT}$. Eventually Coulomb scattering becomes even more important than Compton scattering and lead to the same result: low energy positrons that have equilibrated with the rest of the cosmic plasma.

\Sfig{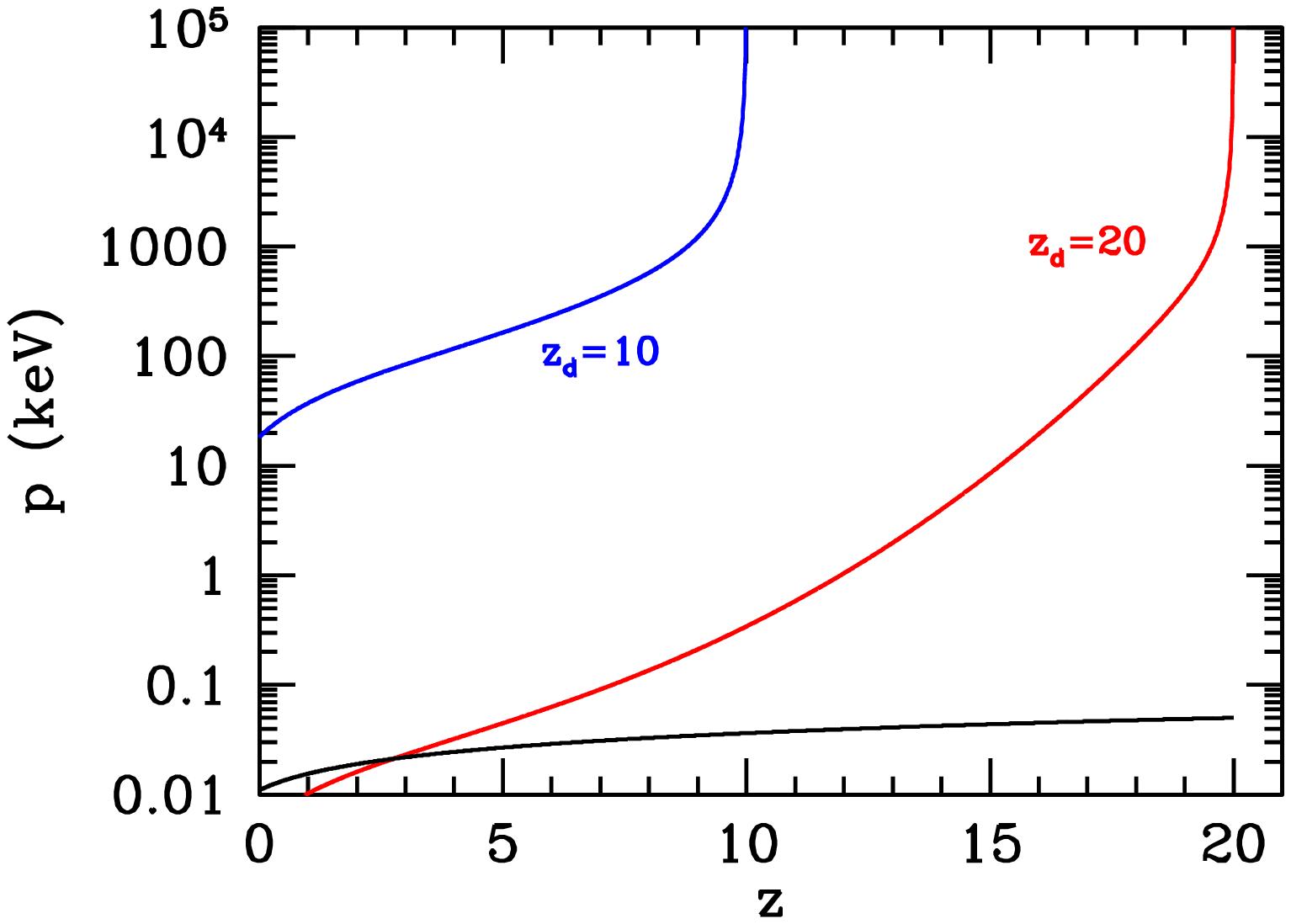}{Momentum of positrons versus redshift for decay redshifts $z_d=20$ (red curve) and $z_d=10$ (blue curve). Result is independent of initial energy $\deltam$. Black curve is $\sqrt{m_e T_\gamma}$, a rough floor on the positron momentum.}

\section{Positron interactions in the galaxy}

Positrons will interact and change energy as they diffuse from the galaxy outskirts to the disk and bulge. As we have seen in the previous section, the relic positrons from dark matter decay thermalize immediately with the CMB and then after structure formation they behave as CHAMPs and are dragged to the galactic halo. Typical velocities for these positrons when they arrive to the disk or the bulge will be of order of 200 km/s. Therefore the kinetic energy of these positrons will be $O (0.1)$ eV.  Positrons and electrons with these energies cannot ionize or excite hydrogen atoms, so their only relevant interactions in the interstellar medium will be elastic Coulomb scattering with  protons or free electrons and annihilation with free or atomic electrons.  The interstellar medium (ISM) is an extremely complex environment \cite{2005pcim.book.....T} because it is made of regions, primarily of hydrogen, with fairly different physical properties. In Table \ref{ism} we present a summary of the different regions in the interstellar medium.  While clouds of molecular ($H_2$), atomic (HI) and ionized Hydrogen (HII regions) occupy a small volume fraction, the intercloud media (Hot Ionized, Warm Ionized and Warm Neutral) pervades most of the volume of the disk and bulge. 

\begin{table}[!b]

\begin{center}
\begin{tabular}{l c c c c c}
\hline
Component &~~Temperature &~ Midplane density & ~Filling fraction  ($\%$)  
 & ~Ionization fraction & ~Scale Height \\
 & (K)& $n_0$ (cm$^{-3}$)& $f$ (\%) & $x_{ion}$ & $H$ (pc)\\

\hline
{\em Clouds} & & & &  \\
H2 Molecular (MM)& 10 - 20  &$10^2-10^6$ & 0.1 & $\lesssim 10^{-4}$& 75 \footnote{Gaussian  scale height $n(z)=f \, n_0  \exp-(z/H)^2$.}\\
HI  Cold Neutral (CNM)& 50 - 100  & 20 - 50 & 2 & $4\times 10^{-4}- 10^{-3}$ & 94 $^a$\\
Traditional HII regions  & 8000 & $1-10^5$& --& $\sim 1$ & 30-100  \footnote{Exponential scale height: $n(z)=f\,  n_0\exp(-z/H)$.}\\
\hline
{\em Interclouds} & & & &  \\
Warm Neutral HI (WNM) & $10^3- 10^4$& $0.2-2$&$\sim 30$ & $0.007-0.05$ & 220 $^a$+ 400 $^b$\\
Warm Ionized HII (WIM) & 8000&$0.1-0.3$&$\sim 20$ & $0.6-0.9$ & 900  $^b$\\
Hot Ionized (HIM) & $\sim 10^6$ & 0.002& $\sim 50$& 1 & 3000 $^a$\\
\hline
\end{tabular}
\end{center}
\caption{Typical parameters of the interstellar medium phases. }
\label{ism}
\end{table}

The complexity of the ISM fades away when the issue is the fate of a positron entering the disk. As we show here, a positron entering any of these regions will very quickly thermalize and then annihilate. To compute the rate of 511 keV flux from these positrons then, we can safely assume that the photons are produced very close to the place where the positron enters the disk/bulge. We develop this argument in this section and then compute the flux of positrons into the disk/bulge -- and therefore the 511 keV flux -- in the next section.

Coulomb scattering between positrons and free electrons is very effective so positrons thermalize very quickly even in regions where the ionized fraction is relatively small. In Appendix \ref{coulomb}, we present the differential cross sections as function of the energy transfer and calculate the thermalization length $R_{\rm therm}$. The rate of positron energy change due to Coulomb scattering with free electrons in a medium with free electron density $n_e$ is given by, 
\bea
\label{eq:Eloss}
\frac{d E}{d t} & =& 
 n_e ~\frac{\alpha^2 }{\beta_e m_e } \log \left( \frac{b_{\rm min}}{b_{\rm max}}\right) \, ,
\eea 
where $\beta_e$ is the typical electron velocity (which depends on temperature) and $b_{\rm min}$ and $b_{\rm max}$ are the minimum and maximum impact parameter. These can be related to the maximal and minimal energy transfer $K_{\rm max}$ and $K_{\rm min}$. In ionized media we can take the maximum impact parameter as $b_{\rm max} = (4 \pi n_e)^{-1/3}\simeq 1 \,{\rm cm}$ for $n_e \sim 1 \,{\rm cm^{-3}}$  and the minimum impact parameter as $b_{\rm min}= 2 \alpha/(m \beta_e^2) \simeq 5 \times 10^{-7}$ cm, which give a Coulomb logarithm of $ \log \left(b_{\rm min}/b_{\rm max}\right) \simeq 15$.
Dividing the typical energy ($3k_BT/2$) by this rate, and then multiplying by the velocity provides an estimate for the thermalization length 
\begin{equation}
R_{\rm therm}\sim  10^{-6} ~{\rm pc} ~\left( {\rm cm}^{-3}\over n_{e^-}\right) \, \left( \frac{T}{10^4 K} \right)^2
\label{thermal}
\end{equation}
a scale that is much smaller than any of the regions listed in Table II. Therefore, no matter where a positron enters the disk or bulge, it will almost immediately thermalize with the ambient medium and then participate in the Galactic rotation of that cloud or inter-cloud medium.


After thermalization, positrons 
will annihilate through one of the four available channels summarized in Table \ref{ann}.  For each annihilation channel $i$ with threshold energy $E_{\rm th}$, the annihilation rate is given by 
\begin{equation}
\Gamma_{\rm ann}(i)\equiv \int^\infty_{E_{\rm th}}\, f_E(E) n_{e,\, H}\,  \sigma_i(E) v(E) {\rm d}E
\end{equation}
where $f_E(E)=2\sqrt{E/\pi (k_B T)^3}~\exp\left[-E/{k_B T}\right]$ is the usual Maxwell-Boltzmann distribution function for energy and $\sigma_i(E)$ and $v(E)$ are the energy-dependent annihilation cross section and velocity respectively. Notice that even though the cross section for charge exchange at temperatures of $T\sim$ eV is suppressed with respect to the two other channels \footnote{The threshold energy for this channel is 6.8 eV.}, the corresponding annihilation rate is by far the strongest. So, even though fewer positrons reach the threshold energy in the thermal distribution, the annihilation rate is still comparable to radiative combination with free electrons i.e. $e^+e^-\to Ps \; \gamma$. The remaining channels occur without Ps formation and are: the well-known direct annihilation channel and annihilation with bound electrons that are one order of magnitude slower. 
Given that Ps formation channels are dominant, we can simply estimate the time scale over which positrons annihilate through Ps formation
to be $\Gamma_{\rm ann}^{-1}\sim 10^{12}$ sec. The thermalized positrons are locked in clouds rotating around the center of the Galaxy with typical speed of order 200 km/sec, so the positron will annihilate before the cloud containing it has traveled 5-10 parsec: effectively the positrons annihilate as soon as they enter the disk/bulge.
From the annihilation rates in Table \ref{ann}, we can expect roughly 5--$10\%$ of the positrons to annihilate without forming Ps, depending on the ionization fraction of the media. This value should be compared with the latest experimental value \cite{Jean:2005af} $f_{Ps}= 0.967 \pm 0.022$. Furthermore, the resulting 511 keV line is broadened due to electrons' thermal velocity by the amount $\sim$1.1 keV $\sqrt{T/10^4~ {\rm K}}$ \cite{Murphy}, consistent with observation. 


\begin{table}[!t]
\begin{center}
\begin{tabular}{l l l  }
\hline
Annihilation Channel & ~~Name~~&~~Annihilation rate \\
 & ~~~~~~ &~~$\Gamma_{\rm ann}(i)$ ~~(sec$^{-1}$) \\
\hline

$e^+H\to Ps +  H^+$&   Charge exchange (ce)&  $~~10^{-12} ~n_H$\\

$e^+e^-\to Ps \; \gamma$& Radiative combination (rc)&  $~~10^{-12}~ n_e$\\

$e^+e^-\to 2\gamma$& Direct annihilation with free $e^-$ (daf) & $~~2\times 10^{-13} ~n_e$ \\

$e^+H\to 2\gamma +  H^+$&   Direct annihilation with bound $e^ -$ (dab)&  $~~7\times 10^{-14}~ n_H$\\
\hline

\end{tabular}
\caption{ Annihilation channels for a thermal distribution of positrons with $T\simeq$ 1 eV. Values are taken from  \cite{1979ApJ...228..928B}.}
\label{ann}
\end{center}
\end{table}

To go further, we need to understand the extent of the disk and bulge. 
The interstellar medium (ISM), and in particular the WIM, is rather well-known close to the galactic plane \cite{Ferriere:2001rg,Ferriere:2007yq}, however, it is much more uncertain at more than 1 or 2 kpc above the disk \cite{Gaensler:2008ec}. Therefore, in this work we will adopt a simplified model where we take the interstellar medium  to be a cylinder of radius $R_1$, allowing $R_1$ to vary between $5~ {\rm kpc} < R_1 < 15~{\rm kpc}$, and height $2 h$ around the galactic center. Taking into account the small thermalization radius compared with typical size of the clouds, we do not need a precise description of the different regions inside this cylinder. 
The final ingredient in our model is the bulge which we take as an sphere of $r_b \simeq 1.5$ kpc in the center of the galaxy. As we have seen, all positrons arriving to the ISM cylinder or the bulge will get trapped and annihilate close of the boundary of the considered region. Most of the positrons annihilating in the ISM cylinder will produce 511 keV photons in the disk, while in the inner 1.5
kpc, for $h \lsim 1.5$ kpc, the produced photons will be seen as originating from the bulge. 

\section{Flux of positrons in the galaxy}

As we have seen in the previous section, in practice, all the positrons entering the ISM cylinder or the bulge get trapped and annihilate the first time they cross the galaxy. Here we compute this incoming positron flux and turn it into a 511 keV flux from both the bulge and the disk.
In both cases, the number density of positrons is governed by the rate equation:
\bea
\label{Eq:npos}
\frac{d n_+}{d t} &=& - n_+ \Gamma_{\rm ann} +  S\, ,
\eea
where $S$ is the flux of positrons into the region (either bulge or disk) and $\Gamma_{\rm ann}$ is the annihilation rate. Since the annihilation rate is very large, the two terms on the right cancel and the number of positrons reaches equilibrium at
\bea
\label{Eq:eqnpos}
n_+ &=& \frac{S}{\Gamma_{\rm ann}}.
\eea
To obtain the luminosity of 511 keV photons from this positron density we need to multiply by the annihilation rate, thereby canceling the denominator, and a factor of $2(1-\frac34 f_{Ps})$ to account for the two photons produced and positronium formation. 
The total flux from a given region is then the volume of that region $d^3x$ weighted by $2(1-\frac34 f_{Ps})S/(4\pi d^2)$ where $d$ is the distance of the region from us. We expect then a flux of 511 keV photons from incoming positrons equal to
\be
F = \frac{1-\frac34 f_{Ps}}{2\pi} \int\,d^3x \frac{S(\vec x)}{d(\vec x)^2}
.\ee
In our simplified model, the annihilations take place on the surface of the region (bulge or disk) so $S$ has a delta function restricting the 3D integral
over $d^3x$ to the 2D surface $\Sigma$. Other than the delta function, the source term is the flux, which weights the density of positrons by the
the normal component of the velocity, so the flux of 511 keV photons reduces to
\be
F = \frac{1-\frac34 f_{Ps}}{2\pi} \int_{\Sigma} \,dA_{\hat n} \frac{1}{d^2} \, \int\, d^3v\,  f_+(\vec v,\vec x)\, (-\vec v\cdot \hat n) \, \Theta(-\hat v\cdot \hat n)
\eql{flux5}
.\ee
Here, $f_+$ is the occupation number of positrons, which depends on both position and velocity; $\hat n$ is normal to the surface of either the bulge or the disk; and $dA_{\hat n}$ is the differential area traced out by $\hat n$. The final step function $\Theta$ ensures that we count only those positrons entering the disk or bulge, not those leaving.

The distribution of positrons in the halo is the key ingredient needed to evaluate the flux in \ec{flux5}. The simplest first estimate is to assume the positrons trace the dark matter in the halo, with the difference that their density is smaller by a factor of $\omega$, as defined in \ec{defom}. Taking the simplest possible dark matter profile, an isothermal distribution, leads to
\be
n_+(r)= \omega \frac{\sigma^2}{2\pi G r^2 m_\chi}.
\ee
We fix $\sigma$ by requiring the local dark matter density to be equal to $0.3$ GeV/cm$^3$; this translates into $\sigma=117$ km/sec. The corresponding velocity distribution is then
\be
\label{Eq:isotherm}
f_+^{\rm iso}(r,v)= \frac{n_+(r)}{(2 \pi \sigma^2)^{3/2}}~ e^{-v^2/2 \sigma^2}\, .
\ee

In the case of annihilating positrons, there is an additional issue to consider: positrons that have crossed the disk previously no longer exist as they have annihilated, so we need to count only those positrons that are nearing the disk or bulge for the first time now. Again we choose to model the orbits in a simple way to estimate this survival probability. We assume that the dark matter dominates the gravitational potential, which is spherically symmetric, and equal to $\Phi(r) = \sigma^2 \ln(r/r_v)$ with the zero-point offset $r_v$ chosen to be 200 kpc. In this potential orbits are specified by energy $E=v^2/2+\Phi(r)$ and angular momentum $L=\vert \vec r\times\vec v\vert$. Each orbit has two turning points ($r_1,r_2$) given by the two solutions to: $2(E-\Phi)-L^2/r^2=0$, and period
\be
P = 2\int_{r_1}^{r_2} \frac{dr}{\sqrt{2(E-\Phi(r)) - L^2/r^2}}
.\ee
The survival probability is set to one if, for given $E$ and $L$, $r_1$ is smaller than $r$ (where $r$ is the distance to the surface of the disk or bulge) and the period is very large; we choose the requirement $P>8$ Billion years. Otherwise, the survival probability, $s(\vec v,\vec x)$, is set to zero. Together these two constraints amount to the statement that only positrons on highly radial orbits are still available to fuel the 511 keV radiation today. It is straightforward to show numerically that these two constraints translate into separate restrictions on $L$ and $E$. Requiring $r_1<r$ is satisfied as long as
\be
L < 3.3 r\sigma,
\ee
while the long period requirement is satisfied as long as
\be
E > 1.2\sigma^2.
 \ee
With these additions, the distribution function in \ec{flux5} becomes
\be
f_+(r,v=\sqrt{2(E-\Phi(r))},L) =  f_+^{\rm iso}(r,v)  s(E,L)
\ee
with
\be
s(E,L) = \Theta(E-1.2\sigma^2)\, \Theta(3.3r\sigma - L)
.\ee

\subsection{Bulge}

We first compute the 511 keV flux from the bulge. For simplicity we can take all points on the surface of the bulge, at roughly $r_b=1.5$ kpc, to
lie a distance of $R_\odot=8$ kpc from us, so that $d(\vec x)$ comes out of the integral in \ec{flux5}. The integral over the surface has area $r_b^2d\Omega_{\hat n}$ leaving
\be
F_b = \frac{(1-\frac34 f_{Ps})r_b^2}{2\pi R_\odot^2} \int d^2\Omega_{\hat n}  \, \int\, d^3v\,  f_+^{\rm iso}(v,r_b)\,s(E,L)\, (-\vec v\cdot \hat n) \, \Theta(-\hat v\cdot \hat n)
.\ee
In this spherically symmetric case, the integrand does not depend on the position on the 2D sphere, so the $d^2\Omega_{\hat n}$ integral can be done immediately giving a factor of $4\pi$. The remaining integral over the velocity is azimuthally symmetric so reduces to
\be
F_b = \frac{(1-\frac34 f_{Ps})4\pi r_b^2}{R_\odot^2} \int_0^\infty dv \,v^2  f_+^{\rm iso}(v,r_b) \int_{\pi/2}^\pi d\theta\, \sin\theta \,s(E,L)\, (-v\cos\theta) 
\ee
where the step function has been incorporated into the limits on the polar angle $\theta$: only inward orbits, those with angle $\theta$ between the velocity and the radial vector anti-aligned ($\cos\theta<0$), are included.

To implement the constraints imposed by the survival probability, it is simplest to change dummy variables from $(v,\theta)$ to $(E,L)$, where $L=vr_b\sin\theta$. The Jacobian is $v^2r_b\vert\cos\theta\vert$, so dividing by this, the integrals become
\bea
F_b &=& \frac{4(1-\frac34 f_{Ps})\pi}{R_\odot^2} \int_0^{1.2\sigma^2} dE\, f_+^{\rm iso}\left(v=\sqrt{2[E-\Phi(r_b)]},r_b\right) \,\int_0^{3.3r_b\sigma} dL\,L
\vs
&=&
0.275\,(3.3)^2 \left[ 1-e^{-1.2}\right] \frac{r_b}{r_v} \,\frac{\sigma r_b^2 n_+(r_b)}{\sqrt{2\pi} R_\odot^2}
.\eea
Plugging in numbers leads to
\be
F_b = 217 \omega \left( \frac{100\,{\rm GeV}}{m_\chi}\right)\, \left(\frac{r_b}{1.5\,{\rm kpc}} \right)\,{\rm cm}^{-2}\, {\rm s}^{-1}
.\ee
This means that in order to match the observation of $10^{-3}~{\rm cm}^{-2} \,{\rm s}^{-1}$, we would need a value of $\omega$, 
\be
\omega \simeq 4.7 \times 10^{-6}~\left(\frac{m_\chi}{100\, {\rm GeV}}\right)
\label{final}.
\ee
\subsection{Disk}

The flux from the disk is more complicated to compute for two reasons: (i) the distance to us now varies significantly as one moves along the surface of the disk, so $d$ cannot be removed from the integral in \ec{flux5} and (ii) the normal $\hat n$ is no longer parallel to the radial vector that enters the definition of the angular momentum. Whereas for the bulge $\vec v\times\vec r=vr\sin\theta$ and $\vec v \cdot \hat n=v\cos\theta$, for the disk these two products involve different angles.  Specifically, if we choose the polar axis of the $d^3v$ integral to lie parallel to the radial vector $\hat r$, then $L$ will still be equal to $vr\sin\theta$, but now
\be
\vec v\cdot \hat n = \frac{v}{r}\left[ R\sin\theta\cos\phi + h\cos\theta\right]
.\ee
 Here the 3D distance $r=\sqrt{R^2+h^2}$ where $R$ is the cylindrical radius from the center of the Galaxy; $h$ is the height of the disk; and $\phi$ is the azimuthal angle in the velocity integration. Thus the flux coming from the disk is
 \bea
 F_d &=& \frac{2(1-\frac34 f_{Ps})}{2\pi} \int_0^{R_d} \frac{dR\,R}{\sqrt{R^2+h^2}} \, \int_0^{2\pi} \frac{d\alpha}{ R^2+h^2+R_\odot^2-2RR_\odot\cos\alpha} \\
 &\times&
 \int_0^\infty dv\,v^3f_+^{\rm iso}(r,v) \int_0^\pi d\theta\,\sin\theta\,  s(E,L) \int_0^{2\pi} d\phi \left[ -R\sin\theta\cos\phi - h\cos\theta\right]\, 
 \Theta\left[ -R\sin\theta\cos\phi - h\cos\theta\right] \nn
.\eea
We can simplify this equation in the limit $h \ll R$, which is approximately true
in our simplified model wherein the radius of the disk is much larger than its height,
although we keep $h$ in the denominator to avoid the singularity in $R = R_\odot$. So,
 \bea
 F_d &\simeq& \frac{2(1-\frac34 f_{Ps})}{2\pi} \int_0^{R_d} \,dR \, \int_0^{2\pi} \frac{d\alpha}{ R^2+h^2+R_\odot^2-2RR_\odot\cos\alpha} \vs
 &\times&
 \int_0^\infty dv\,v^3f_+^{\rm iso}(r,v) \int_0^\pi d\theta\,\sin\theta\,  s(E,L) \int_0^{2\pi} d\phi \left[ -R\sin\theta\cos\phi \right]\, 
 \Theta\left[ -R\sin\theta\cos\phi \right]
.\eea
Now $\Theta\left[ -R\sin\theta\cos\phi \right]$ selects $\pi/2 < \phi < 3 \pi/2$ and the integral on $\phi$ is trivial, becoming just $2R\sin\theta$. To proceed in this case it is more convenient to change the $v$ integral to $K \equiv v^2/(2 \sigma^2)$. Then, the restriction on $L= v r \sin \theta < 3.3 r \sigma$ can be set on K and we have, $1.2  - \Phi(r) /\sigma^2< K < (3.3/\sin \theta)^2/2$.  The upper bound on the $K$ integral must be larger than the lower bound, so $\sin^2
\theta < 3.3^2 / [2 (1.2  - \Phi(r)/\sigma^2)]$. For large $r$, $\Phi$ is sufficiently large that this constraint is always satisfied. But for $r < 3 $ kpc we have a bound on $\sin^2\theta$. Expressing the integral as one over $\cos\theta$ then leads to
\bea
F_d &\simeq& \frac{2(1-\frac34 f_{Ps})}{2\pi} \int_0^{R_d} dR\,R \, \int_0^{2\pi} \frac{d\alpha}{ R^2+h^2+R_\odot^2-2RR_\odot\cos\alpha}~ 4~ \frac{n_+(R) \sigma}{ (2 \pi )^{3/2}}\vs
 &\times&  \left[ \int_{a_r}^{1}~ d \cos\theta
 \sin \theta ~ \int_{1.2 - \Phi(R)/\sigma^2}^{3.3^2/(2\sin^2 \theta)} K ~ dK~ e^{-K} 
+ \int^{-a_r}_{-1}~ d \cos\theta  \sin \theta ~ \int_{1.2 - \Phi(R)/\sigma^2}^{3.3^2/(2\sin^2 \theta)} K ~ dK~ e^{-K} \right]
\eql{fdmid}
\eea
where $a_r$ is defined as
\be
 a_r^2 = \begin{cases}
 1 - \frac{3.3^2}{2 (1.2  - \Phi(r)/\sigma^2)}& r<3\, \text{kpc}\cr
 0 & r>3 \, \text{kpc}
 \end{cases}
 .\ee 
The dimensionless factor inside the square brackets in \ec{fdmid} can be evaluated numerically; a good fit is 
 $C_2(R)\simeq 0.007 + 0.01 (\frac{R}{1 ~{\rm kpc}} -1)$ for $R$ from 1 to 20 kpc (again we neglect the small distinction here between $r$ and $R$). Then we have:
\bea
F_d &\simeq& \frac{2(1-\frac34 f_{Ps})}{2\pi} \int_0^{R_d} dR\,R \, \frac{2 \pi }{\sqrt{ (R^2+h^2+R_\odot^2)^2-4R^2R_\odot^2}}~ 4~ \frac{n_+(R) \sigma}{ (2 \pi )^{3/2}} ~ C_2(R)
\eea
From here, we obtain an 
estimate of $(B/D)^{-1}$ by dividing by the flux obtained from the bulge:
\bea 
\left(\frac{B}{D}\right)^{-1}(R_d) &=&\frac{8}{2 \pi} \int_1^{R_d} d R 
\frac{C_{2}(R)~ R~Ṛ_\odot^2}{C_b ~r_b^2~ \sqrt{(R^2 + R_\odot^2+h^2)^2 - 4 R^2 R_\odot^2}} \frac{n_+(R)}{n_+(r_b)}
\eea
with $C_b=(3.3)^2 \left[ 1-e^{-1.2}\right] \frac{r_b}{r_v}\simeq 0.057$. Integrating this expression numerically, we obtain the value of $B/D$. Given the crudeness of our disk model, we compute this for a variety of values of $R_d$, the radius of the disk.The result is shown in Figure \ref{fig:BoDiso}. \Sfig{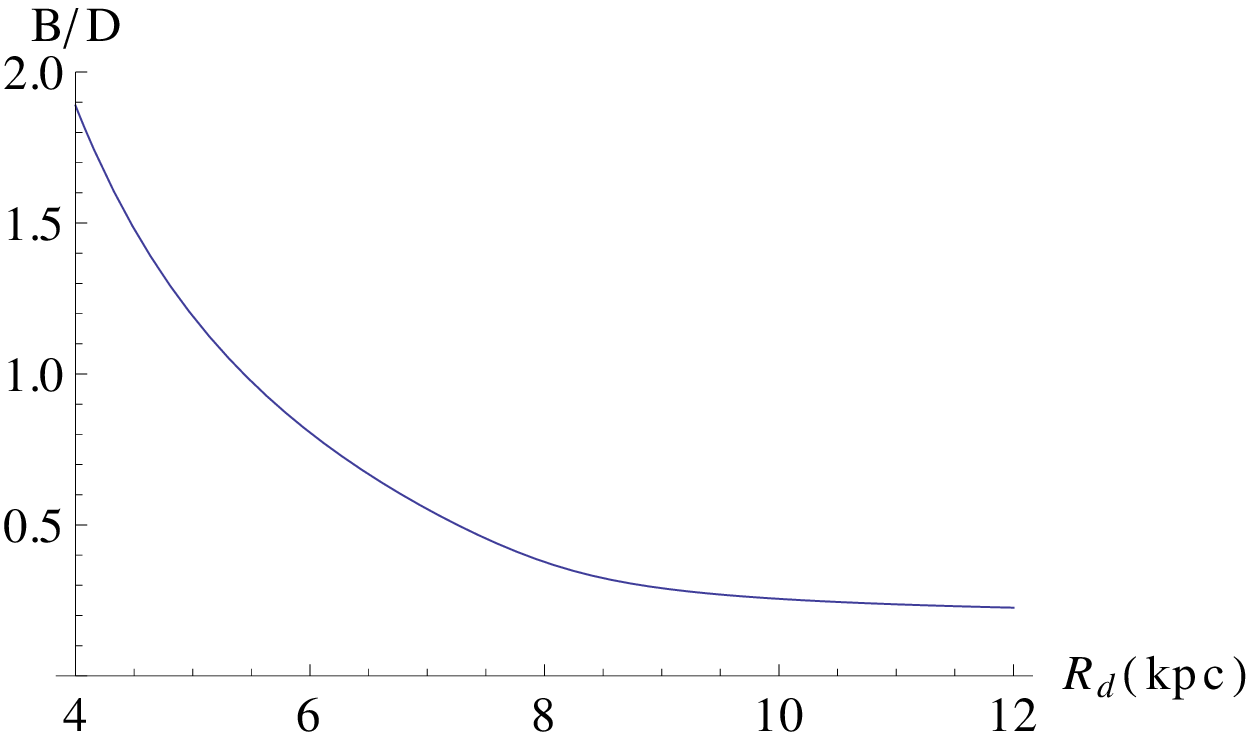}{Ratio $B/D$ of the flux of 511 photons from the bulge and the disk for different values of the ISM disk radius $R_d$ as explained in the text.}
For the simple isothermal profile, the bulge to disk ratio is of order unity (in agreement with observations~\cite{Weidenspointner:2007rs}) if the disk scale length is less than about 8 kpc, but drops to 0.25 for larger values of the disk radius.


\section{Discussion}

High energy positrons produced at early times via the decay of a second dark matter species cool down and will eventually get trapped in galaxies. These cooled positrons would contribute to the 511 keV flux in our Galaxy and -- with a suitable choice of parameters -- might explain the observed bulge to disk ratio of this flux. 
We conclude with a brief discussion of the possible models that can generate this kind of mass hierarchy and couplings in the dark sector.

%

First consider the case where the unstable DM component is neutral. In supersymmetric models, while the possibility of two neutral species is quite natural (with both a neutralino and gravitino), the heavier species will often decay predominantly into photons, leading to very tight constraints from the diffuse flux~\cite{Boubekeur:2010nt}. This traces back to the fact that the lightest neutralino usually has a large photino component. 
More generally, we can use effective field theory to write down operators that lead to decays. If the dark matter species consists of fermions, the coefficient is of order $\Lambda^{-2}$ where $\Lambda$ is the UV scale above which the effective theory breaks down. The long lifetimes required then point to $\Lambda\sim10^{11}$ GeV. For scalar dark matter, the operator is suppressed by only a single power of $\Lambda$, so the long lifetimes require a UV scale of order the Planck mass. In this effective field theory context, too, operators leading to decay to photons must be suppressed. Models that accomplish this typically rely on a secluded dark sector, which communicates with the Standard Model through suppressed interactions. In all models, it is a challenge to obtain the correct relic density since couplings are so small. Likely, $\chi$ cannot be a conventional thermal relic and other mechanisms (such as `freeze-in''  \cite{Hall:2009bx} or dilution via a short phase of thermal inflation \cite{Lyth:1995hj}) need to be explored.

Getting the correct relic density is even more difficult in the case of charged dark matter. In the charge symmetric case, dark matter is likely to annihilate quickly into photons, leaving no appreciable $X$'s to decay at late times. So a mechanism to suppress annihilations will be essential to any successful symmetric model. Asymmetric models do not suffer from this problem, but generating an asymmetry -- which of course requires out-of-equilibrium -- might prove challenging in a sector that interacts electromagnetically.


Although we have touched on particle physics realizations of our scenario, in principle our constraints apply to any source of positrons that are operating from recombination up to now. Most of the constraints (apart from heavy water, colliders, and possibly direct decays to photons) apply to any mechanism that produces positrons in the early universe.

\acknowledgments
We thank N. Gnedin, D. Hooper, A. Konigl, A. Kravtsov, A. Santamaria and Y. Ascasibar  for useful discussions. L.B. and O.V.  acknowledge  partial support by MEC and FEDER (EC), Grants No. FPA2008-02878 and FPA2011-23596 and by 
the Generalitat Valenciana under the grant PROMETEO/2008/004. S.D. is supported by the U.S.
Department of Energy, including grant DE-FG02-95ER40896, and by the National Science
Foundation under Grant AST-090872.

\bibliography{positron}

\appendix
\section{Diffuse Spectrum from Inverse Compton Scattering}

When high energy positrons are produced, they scatter down off of the CMB, up-scattering the photons to energies where they redshift freely and can be observed today as part of the diffuse background. Here we calculate the diffuse spectrum from this process of inverse Compton scattering. At any given time, an energy density $\Delta\rho = \deltam n_X dt/\tau$ is deposited into diffuse photons, where $\deltam$ is the energy carried by electrons and positrons in a single $X$ decay (roughly equal to $m_X/2$ if the decay is 2-body, the branching ratio to positrons is unity, and $m_X\gg m_\chi$). The spectrum of photons produced by inverse Compton scattering scales as $d\rho/dE\propto E^{-1/2}$ \cite{Dimopoulos:1988zz} up to a maximum energy of order $E_{\rm max}=4T(\deltam/m_e)^2$, so
the differential spectrum produced by decays in a small interval $dt$ is
\be 
\frac{d^2\rho}{dtdE} = \frac{\deltam n_X}{2E_{\rm max}^{1/2}E^{1/2}\tau} \Theta\left[ E_{\rm max} - E\right]
.\ee
Note that, since $E_{\rm max}^{1/2}$ scales as $\deltam$, the factors of $\deltam$ in the amplitude cancel out and the differential spectrum depends on $\deltam$ only indirectly in the cut-off energy. This reflects the physical fact that a larger $\deltam$ leads to more energy injected into the diffuse spectrum per decay, but this greater energy extends to larger maximum energy, so that the amplitude of the spectrum at any point below the maximum does not depend on $\deltam$.

To convert this to a spectrum today, use the facts that the energy today $E_0$ is equal to $E/(1+z)$ and the time interval $dt$ is equal to $-dz/[H(z)(1+z)]$, and integrate over all time, remembering that the left hand side scales as $(1+z)^3$. So,
\be 
\frac{d\rho}{dE_0} = \frac{1}{E_0^{1/2}} \int_0^\infty \frac{dz}{H(z) (1+z)^{9/2}} 
\frac{\deltam n_X(z)}{2E_{\rm max}^{1/2}\tau} \Theta\left[ E_{\rm max} - E_0(1+z)\right].
\ee
The argument of the $\Theta$ function is independent of $z$ because $E_{\rm max}$ scales as $1+z$, reflecting the fact that, at any given time, the photons with maximum energy produced by inverse Compton scattering redshift down to the same maximum value of $E_0$ today ($4T_0(\deltam/m_e)^2$). Dividing by the energy, we obtain the spectrum today:
\be 
\frac{dn}{dE_0} = \frac{m_e n_X(t=\tau)}{4T_0^{1/2} \tau E_0^{3/2} H_0 (1+z_d)^3} \Theta\left[ \frac{4T_0(\deltam)^2}{m_e^2} - E_0\right]
I(\tau),
\ee
where 
\be 
1+z_d\equiv 1+z(t=\tau) \simeq 68 \left( \frac{10^{15} {\rm sec}}{\tau} \right)^{2/3}
\ee
the approximate equality holding at the percent level for $1<z<200$, and 
\be 
I(\tau)\equiv \int_0^\infty dz
\frac{e^{-[t-\tau]/\tau}}{(H/H_0) (1+z)^{2}}  
\simeq 9.3\times 10^{-5} \left( \frac{\tau}{10^{15}\,{\rm sec}} \right)^{5/3},
\ee
with the approximate equality valid for $2<z<600$. The spectrum then scales as $E_0^{-1/2}$ with an amplitude governed by the dimensionless integral over redshift, which depends only on the lifetime of $X$.
Plugging numbers and dividing by $4\pi$ sr leads to the present day flux in \ec{textflux}. 

\begin{figure}[t]
\begin{center}
\includegraphics[width=0.7\columnwidth]{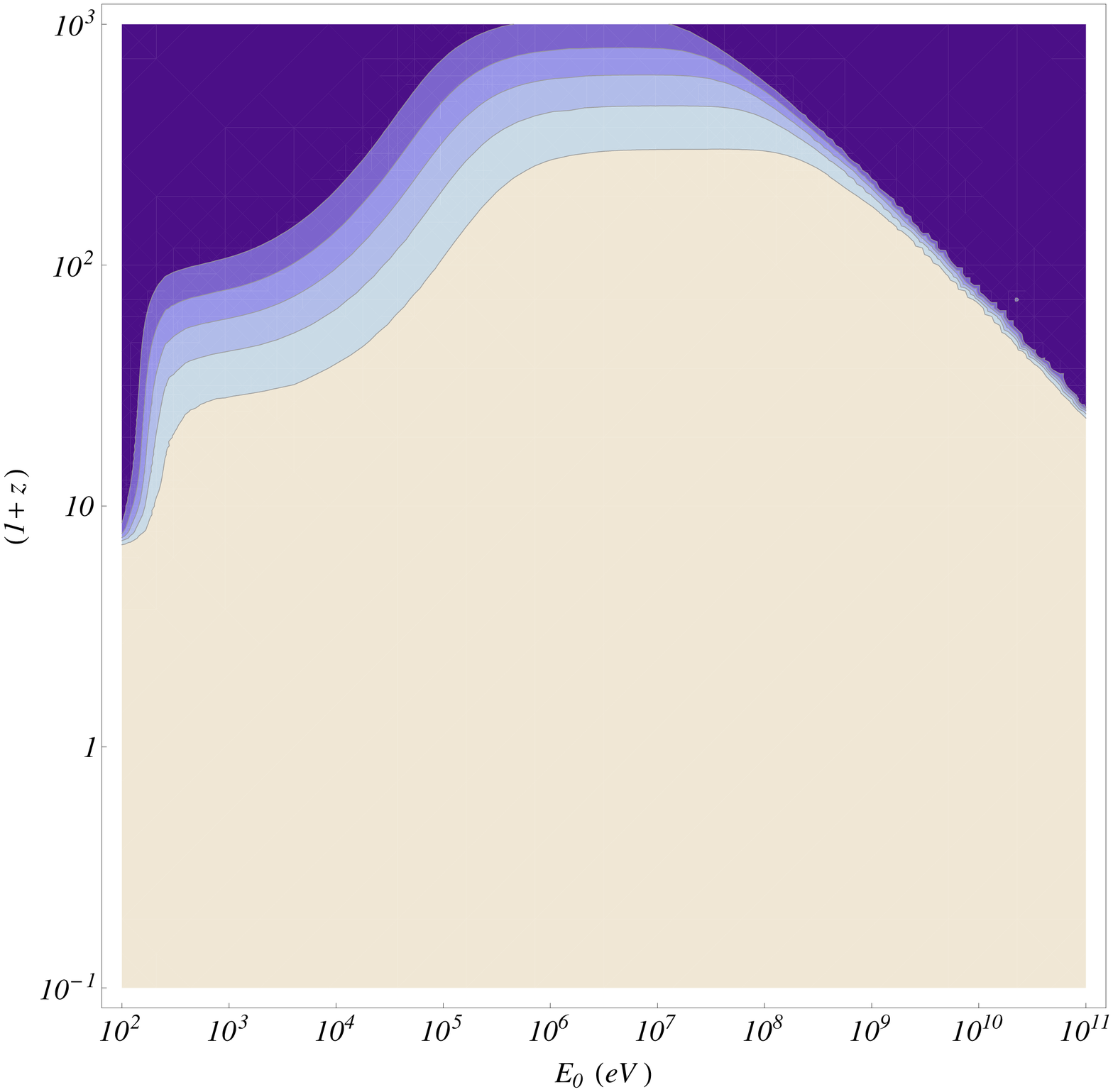}
\caption[text]{ \small Opacity of the universe for different redshifts, $(1+z)$, and photon energies, $E_0$. The different constours correspond to attenuation factors, $e^{-\tau(E_0, z)} = 0.5,~ 0.4,~0.3,~0.2,~0.1, ~0.01$ from light to dark.}
\label{figOD}
\end{center}
\end{figure}

A lingering question is whether the photons produced at large $z$ ftravel freely through the universe or if they interact and are absorbed before arriving here. To answer this question we need to know the optical depths of the absorbtion and scattering processes. This problem was discussed in detail in Refs.\cite{zdziarski88,zdziarski89}. Using the code developed in \cite{Cirelli:2010xx}, we find attenuation as depicted in Fig.~\ref{figOD}.
For most of the parameter space we are interested in here, attenuation is irrelevant, although we do include it in the constraints shown in Fig.~\rf{hw}.


\section{Diffuse spectrum from radiative decays and Final State radiation}

In this appendix, we compute the diffuse spectrum of photons produced through decay of the NLSP (the $X$ particles)~\cite{Fortin:2009rq}. There are two such processes that we will consider here. The first one is radiative decays wherein photons are produced directly. The second process is final state radiation, where a photon is emitted from the produced electron/positron through Bremsstrahlung .
Let us consider a decay process with partial decay width $\Gamma_\gamma$. Each decay leads to $N_\gamma$ photons. For instance $N_\gamma$= 1 for $X\to \gamma\chi$. Following Ref.~\cite{Chen:2003gz}, we write the measured diffuse flux at redshift $z$ as
\be
E_\gamma {d\Phi_\gamma\over E_\gamma} (z)={c\over 4 \pi} \int_z^\infty {dz' \over H(z') (1+z')}\frac{J\({1+z'\over 1+z} E, z'\)}{(1+z')^ 3/(1+z)^3}
\ee
where $H(z)=H_0\sqrt{\Omega_\Lambda+\Omega_M (1+z)^3}$ is the Hubble expansion rate and the emissivity $J(E_\gamma, z)$ takes the form
\be
J(E, \, z)= N_\gamma\, n_X(z)\, E {d\Gamma_\gamma\over d E}
\ee
where $n_X(z)$ is the number density of the decaying particle that can be written as
$$n_X(z)=\omega {\rho_c \, \Omega_{\rm cdm}\over m_X}(1+z)^3 e^{-t(z)/\tau}$$
The present photon flux from a general decay to photons can be written as
\bea
\label{flux}
\frac{d\Phi_\gamma}{dE_\gamma}= \omega N_\gamma \frac{c}{4\pi} \frac{\rho_{c}\, \Omega_{\rm cdm}}{m_X\, E_\gamma} \int^{\infty}_{0} {dz' \over H(z')(1+z')}\, \, e^{-t(z' )/\tau} \; E\, {d \Gamma_\gamma \over d E},
\eea
where $E=E(z')= (1+z')E_\gamma$ is the energy of the photon at production.
\subsection{2-body decays}
If the NLSP decays through a 2 body process $X\to \chi \gamma$, then $N_\gamma=1$ and $d \Gamma_\gamma/ d E = {\rm Br}_\gamma \, \Gamma_X\delta(E_\gamma (1+z')- \epsilon_\gamma)$, where ${\rm Br}_\gamma\equiv \Gamma_\gamma/\Gamma_X$ is the branching ratio to photons, $\epsilon_\gamma={m_X \over 2}\(1-(m_\chi/m_X)^2\)$ is the photon energy at production and $\Gamma_X=\tau_X^{-1}$ is the total decay rate. The delta function can be integrated using the formula $\delta\[f(z)\]=\delta(z-z_\star)/|f'(z=z_\star)|$ where $z_\star$ is the solution which satisfies $f(z_\star)=0$. In our case $z_\star=\epsilon_\gamma /E_\gamma -1$. After integration we get
\bea
\frac{d\Phi_\gamma}{dE_\gamma}=\omega \, \frac{c}{4\pi}\frac{\rho_{c}\, \Omega_{\rm cdm}\, {\rm Br}_{\gamma}}{m_X\tau_X}\frac{e^{-t(z_\star)/\tau_X}}
{E_\gamma H_0 \sqrt{\Omega_\Lambda +\Omega_M (\epsilon_\gamma / E_\gamma)^3}} \Theta(\epsilon_\gamma -E_\gamma),
\eea
where the $\Theta$ function simply cuts energies larger than the initial energy $\epsilon_\gamma$ and $t(z)$ is given by~\cite{Cembranos:2007fj}
\bea
t(z)\equiv \frac{2\log [(\sqrt{\Omega_\Lambda (1+z)^{-3}} +\sqrt{ \Omega_M +\Omega_\Lambda (1+z)^{-3}})/\sqrt{\Omega_M} ]}{3H_0 \sqrt{\Omega_\Lambda}},
\eea
Taking into account that $c/H_0\simeq 1.28\times 10^{28} \cm$, $\rho_{c}=5.46\times10^{-6} ~\mbox{GeV}/\mbox{cm}^3$, $H_0=72 \, \mbox{km}~\mbox{sec}^{-1}~ \mbox{Mpc}^{-1}$, $\Omega_\Lambda=0.7$, $\Omega_{M}=\Omega_{\rm cdm}+\Omega_B=0.3$ and $\Omega_{\rm cdm}=0.25$, we find that the flux is
\bea
\frac{d\Phi_\gamma}{dE_\gamma}&=&
1.12 \times 10^{6}\, {\rm cm}^{-2}\,{\rm sr}^{-1}\,\sec^{-1}\,\mbox{GeV}^{-1} \\ &&\times\,
\omega\, {\rm Br}_{\gamma} \( \frac{\gev}{E_\gamma}\)\(\frac{100\gev}{m_X}\) \(\frac{10^{13}\sec}{\tau_X}\)
\frac{e^{-t(z_\star)/\tau_X}}
{\sqrt{0.7+0.3(\epsilon_\gamma/E_\gamma)^3 }} \Theta(\epsilon_\gamma -E_\gamma)\, .
\eea
%
%
This differential flux of photons must be compared to the observed diffuse gamma ray flux. The maximum of this flux must be smaller than that observed flux by SPI \cite{Churazov}, COMPTEL \cite{Weidenspointner} and EGRET \cite{Strong:2004ry}.
For a given lifetime and $\Delta M = m_{X} - m_{\chi}$, we can find the constraint on the product $\omega~ {\rm Br}_{\gamma}$ as a function of the lifetime $\tau_X$. 

\begin{figure}[h!]
\begin{center}
\includegraphics[]{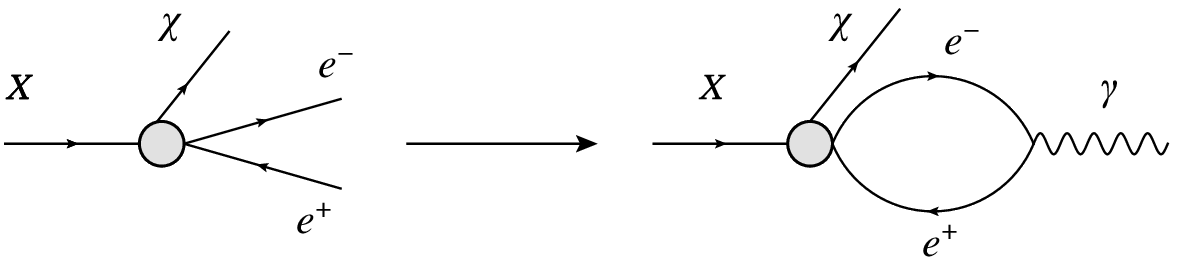}
\caption[text]{ \small Diagram contributing to photon emission at 1-loop for a neutral NLSP. }
\label{fig}
\end{center}
\end{figure}

As an example of 2-body decays leading to a photon,  consider the decay represented in Fig.~(\ref{fig}). Indeed, even though, by assumption, there is no interaction that would mediate the decay $X\to\chi\gamma$ at tree-level, there will be subdominant diagrams leading to photons \footnote{Notice that if the four-fermion interaction leading to the decay $X\to \chi e \bar{e}$ contains a $\gamma_5$, this process will be absent \cite{Kopp:2009et}.} . The first of such diagrams is the one shown in the Figure (\ref{fig}), where one just joins the positron and electron to form a loop, where a photon is emitted. Notice that this decay is {\it 2-body}, so the photon will be {\it monochromatic}. The branching ratio to photons can be estimated as ${\rm Br}_\gamma\simeq {\alpha/ 4 \pi}$.

\subsection{Final State Radiation (FSR)}
\begin{figure}[t!]
\begin{center}
\includegraphics[width=0.7\columnwidth]{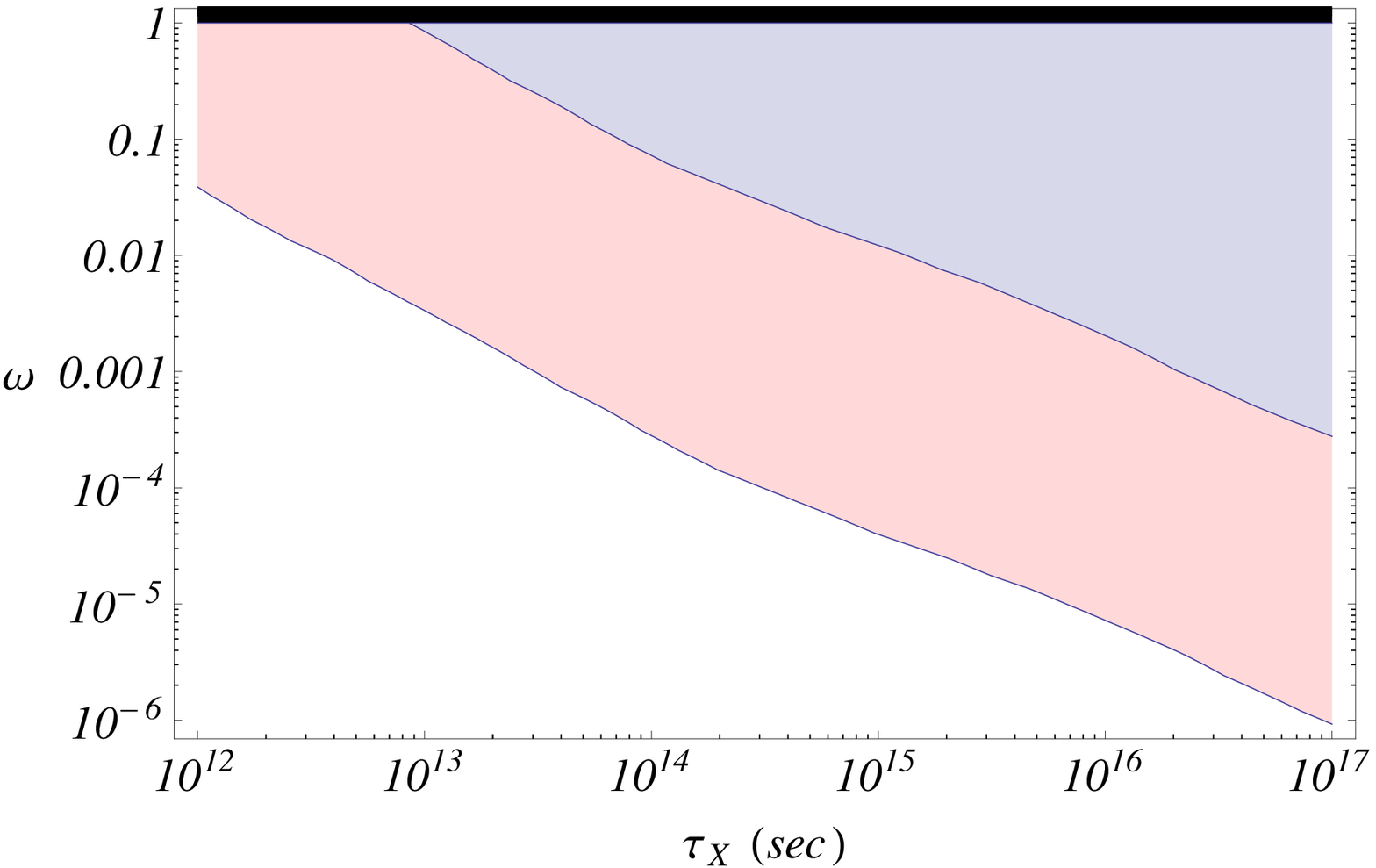}
\caption[text]{ \small Combined bounds on the diffuse flux for $E_i=1\gev$ and $m_X=100\gev$. The pink area is excluded by Brehmsstrahlung emission while the light blue area is excluded by direct decay to photons at one-loop (Fig. (\ref{fig})). 
}
\label{fig5}
\end{center}
\end{figure}

In addition to direct decays to photons, for any dark matter component decaying to a positron or a positron-electron pair, there are processes where photons are emitted from the final state positrons/electrons through Bremsstrahlung. The produced photons, which are called final state radiation (FSR) or internal Bremsstrahlung, will also contribute to the diffuse background of radiation. In the colinear limit, the spectrum of such photons at production is given by \cite{Birkedal:2005ep}
\bea
{d\Gamma_\gamma \over d x}\simeq {\alpha\over \pi} \( 1+(1-x)^2 \over x\)\log \[ {m^2_X\over m^2_e} (1-x)\] \,\Gamma_X
\label{brehm}
\eea
where $x\equiv 2 E/m_X$. In the limit of small mass splitting $x=2 \Delta M/m_X \ll 1$, we get from Eq. (\ref{brehm})
\bea
E {d \Gamma_\gamma \over d E}\simeq {4 \alpha \over \pi} \log\( m_X\over m_e \) \Gamma_X \, .
\eea
Plugging this expression in Eq. (\ref{flux}), we get the diffuse flux from internal Brehmsstrahlung
\bea
\frac{d\Phi_\gamma}{dE_\gamma}={\omega \,\alpha \over E_\gamma} \frac{c}{\pi^2}
\frac{\rho_{c}\, \Omega_{\rm cdm}}{m_X} \log\(m_X\over m_e\) \Theta(\epsilon_\gamma -E_\gamma)\, ,
\eea
where we have used
\bea
\int^\infty_0 dz\, \frac{e^{-t(z)/\tau_X}}{(1+z)\sqrt{\Omega_\Lambda +\Omega_M (1+z)^{3}}} = {\tau_X \, H_0}.
\eea
We can represent both bounds from direct decay and FSR in Fig (\ref{fig5}); note that these bounds are consistent with those obtained in Refs.~\cite{Beacom:2004pe,Beacom:2005qv}.

\section{Coulomb energy losses}
\label{coulomb}
We consider here the problem of elastic Coulomb scattering of positrons on interstellar matter. The positron kinetic energy corresponding to a velocity of 200 km/s is of order $\sim 0.1$ eV so is too low to ionize or excite neutral Hydrogen. Thus, at these low energies, positrons will scatter-off only free electrons or protons. Since the protons are very heavy, the energy transfer is completely negligible. Therefore, the energy transfer is dominated by Coulomb scattering of positrons and free electrons. 

Consider the elastic scattering of an electron and positron. Following Jackson \cite{1998clel.book.....J}, we work in the the $e^-$ rest frame and translate results to  the Lab frame.  The scattering is represented in  Fig.~\rf{coulomb2}.   The variable defined in the figure are $b$ the impact parameter; $\psi$ the incident angle, and $\theta$ the scattering angle.   
When passing through its trajectory, the electron traveling at a velocity $\beta$ will feel the electrostatic attraction given by Coulomb's law $\vec{F}={\alpha \over r^3} \vec{r} $. Only the force perpendicular to the trajectory will give an effect, as the parallel one cancels out, so
 the momentum transfer is 
$$
\Delta p= \int^\infty_{0} F_\perp\, \d t=-{\alpha\over b \beta}\int_0^\pi \sin\psi\, \d\psi = -{2 \alpha\over b\beta}
$$
where we have changed variable using $\d t= \beta \d x$ where $\d x=-b \,\d\theta/\sin^2\theta$.  
\Sfig{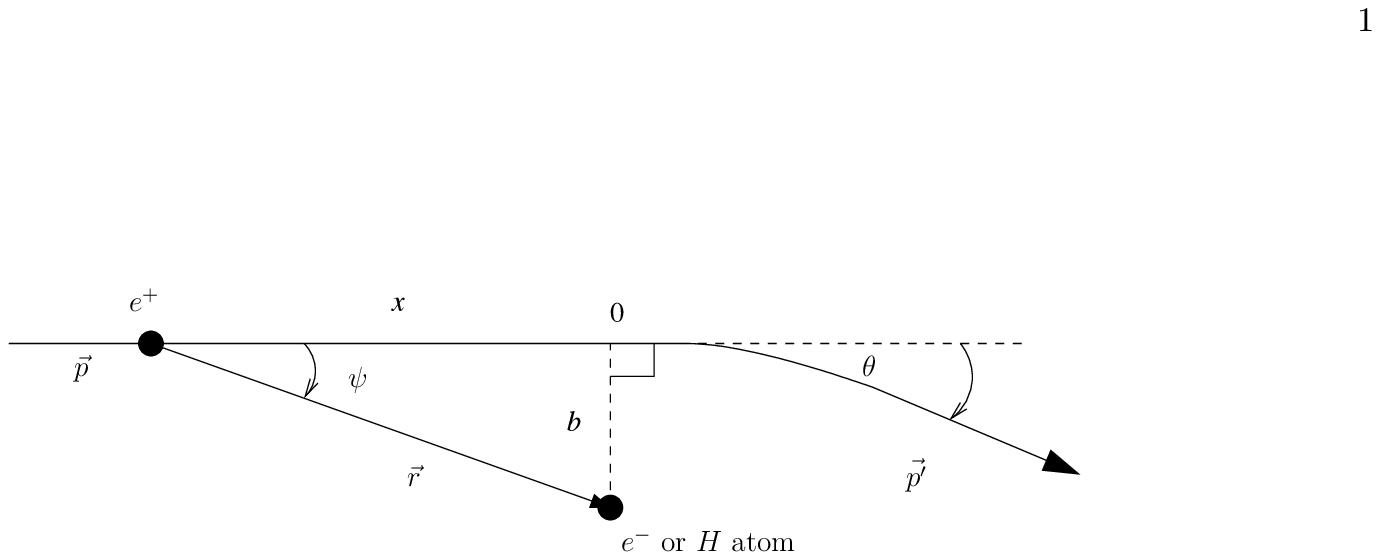}{The geometry of Coulomb scattering.}

The energy transfer \emph{for each scattering} is given by 
\begin{equation}
\Delta E = {(\Delta p)^2 \over 2 m_e}={2 \alpha^2 \over b^2 \beta^2 m_e}\,.
\label{deltaeb}
\end{equation}
In order to obtain the total energy transfer for multiple collisions, we have to integrate over the number of electrons contained in the cylinder with volume element $2 \pi b\, \d b\,  \d x$. Since $dx=\beta dt$, we have the energy loss rate
\bea
{\d E\over \d t}=-\Delta E\,  \frac{\beta n_e}{4 \pi} (2 \pi b\, \d b)\,
\eea
where we divided by $4\pi$ to average over all directions.
Integration over the impact parameter leads to
\begin{equation}
-{\d E\over \d t}{\Big |}_C={\alpha^2\over \beta m_e}n_e
\log\(\frac{b_{\rm max}}{b_{\rm min}}\)\, , 
\label{eloss}
\end{equation}
where $b_{\rm max}$ and $b_{\rm min}$ are the maximum and minimum impact parameters respectively. The energy loss depends only logarithmically on the cutoffs so it is sufficient to obtain rough estimates of $b_{\rm max}$ and $b_{\rm min}$. The minimum impact parameter corresponds to the closest approach and thus to the maximum energy transfer, so  $b_{\rm min}= {2\alpha\over m\beta^2}$. It is more difficult to define define $b_{\rm max}$ for free electrons due the the infinite range of the electromagnetic force. For any given impact parameter, each scattering should take place in a volume $\Delta V(b) =2 \pi b^2 \beta \Delta\tau$ where $\Delta\tau$ is the scattering duration. There should be one electron in $\Delta V(b)$, so the maximum impact parameter $b_{\rm max}$ should satisfy $n_e \Delta V(b_{\rm max})=1$.  Thus $b_{\rm max}= (4 \pi n_e)^{-1/3}$.
So, the Coulomb logarithm can be written as \cite{2000taap.book.....P} 
\begin{equation}
L_C \equiv \log(b_{\rm max}/b_{\rm min})\sim \log(3 k_B T \lambda_D/2 \alpha)\, \simeq 15, 
\label{lc}
\end{equation}
where $\lambda_D=(k_B T /4 \pi n_e \alpha )^{1/2}$ is the Debye length; the screening length in a fully ionized medium, i.e. a plasma.  

Thus, the positron will reach the temperature of the plasma after a time 
$t_{\rm therm.}$, given by,
 \bea
\label{eq:tterm}
t_{\rm therm.} & =& \frac{3 k_B T/2}{d E/d t} =  \frac{3 k_B T \beta m_e} { 2 n_e~ \alpha^2 L_C}~ \,.
\eea 
This is associated with a thermalization length
\begin{equation}
R_{\rm therm}\equiv \beta t_{\rm therm} \simeq  10^{-6} ~{\rm pc} ~\left( 1/{\rm cm}^{3}\over n_e\right) ~\left(\frac{T}{10^4\,K}\right)^2.
\label{athermal}
\end{equation}

\end{document}